\documentclass[aip,jcp,prreprint,noshowkeys,superscriptaddress]{revtex4-1}

\usepackage[fleqn]{amsmath}
\usepackage{bm}
\usepackage{amssymb}
\usepackage{physics}
\usepackage{graphicx}
\usepackage{float}
\usepackage{hyperref}
\usepackage{mathptmx}
\usepackage[version=4]{mhchem}
\usepackage{siunitx}
\usepackage[miktex]{gnuplottex}
\usepackage{subcaption}

\usepackage{tikz}
\usetikzlibrary{tikzmark,calc}

\usepackage{algorithm}
\usepackage{algpseudocode}

\usepackage{natbib}
\AtBeginDocument{\nocite{achemso-control}}

\usepackage{xcolor}
\usepackage{ulem}
\definecolor{stijncolor}{cmyk}{0,0.5,1,0} 

\newcommand{\su}{\uparrow}
\newcommand{\sd}{\downarrow}

\newcommand{\UCAM}{Yusuf Hamied Department of Chemistry, University of Cambridge, Lensfield Road, Cambridge, CB2 1EW, U.K.}
\newcommand{\UNB}{Department of Chemistry, University of New Brunswick, Fredericton, Canada}
\newcommand{\UNBMATH}{Department of Mathematics \& Statistics, University of New Brunswick, Fredericton, Canada}

\hypersetup{
    colorlinks=true,
    linkcolor=blue,
    filecolor=magenta,      
    urlcolor=cyan,
    citecolor=blue,
    }
\graphicspath{./images/}
\begin{document}
\title{Spin-Symmetry Projected constrained Unrestricted Hartree--Fock}
\author{Amir~Ayati}
\affiliation{\UNB}
\author{Hugh~G.~A.~Burton}
\affiliation{\UCAM}
\author{Stijn~De~Baerdemacker}
\affiliation{\UNB}
\affiliation{\UNBMATH}
\date{\today}
\begin{abstract}
We introduce an electronic structure approach for spin symmetry breaking and restoration from the mean-field level. The spin-projected constrained-unrestricted Hartree--Fock (SPcUHF) method restores the broken spin symmetry inherent in spin-constrained-UHF determinants by employing a non-orthogonal Configuration Interaction (NOCI) projection method. This method includes all possible configurations in spin space compatible with a Clebsch--Gordon recoupling scheme in a NOCI calculation. The tunable one-pair-at-a-time characteristics of the symmetry-breaking process in c-UHF allow us to reduce the computational costs of full projection. SPcUHF is tested on 4-, 6-, and 8-electron systems that exhibit dominant static and/or dynamic correlations.
\end{abstract} 
\maketitle

\raggedbottom

\section{Introduction} 
The goal of electronic structure theory in quantum chemistry is to solve the many-body Schr\"odinger equation, striking a balance between reducing computational cost while maintaining accuracy through effective and physically motivated approximations. A foundational approach for solving the electronic structure problem is through Hartree--Fock (HF) mean-field theory. HF employs and optimizes a single Slater determinant for the wavefunction, accounting for all interactions except electron correlation.\cite{ring2004nuclear} Several post-Hartree-Fock methods have been developed to address this lack of electron correlation. Coupled Cluster theory, \cite{kummel2003,bartlett2007,cramer2003, shavitt2009} for example, is recognized for its precision in recovering dynamic correlation, while Multi-Configurational Self-Consistent Field (MCSCF) theory \cite{wolinski1987,mcdouall1988,nakano1993,roos1980} is better suited for modeling static correlation. Additionally, techniques such as the Density Matrix Renormalization Group (DMRG), \cite{white1999,nishino1995,marti2010,stoudenmire2012} and full configuration interaction Monte-Carlo (FCIQMC), \cite{booth2009fermion} are best suited for static correlation where a configuration interaction is essential. Although computationally intensive, these methods strive for a more accurate representation of electron behavior in complex chemical systems.

The mean-field framework sometimes results in solutions that break essential symmetries of the system, often linked to phase transitions. \cite{vcivzek1967stability, paldus1970stability, fukutome1981unrestricted, calais1985gap} For instance, the Unrestricted-HF (UHF) method shows a breaking of spin symmetry at the Coulson--Fischer (CF) point during bond dissociation,\cite{coulson1949} leading to a more energetically favorable solution in the symmetry-broken sector, associated with static correlation.\cite{szabo2012modern} Similar mechanisms are observed in the Bardeen–Cooper–Schrieffer (BCS) and Hartree–Fock–Bogoliubov (HFB) theories for superconductivity \cite{mayoh2021evidence, bauer2007spectral, muther2002spontaneous} and nuclear superfluidity, \cite{dean:2003,brink:2005} where the associated mean-field wavefunctions do not represent states with a well-defined particle number.\cite{chen2023robust} This breaking of symmetry introduces the onset of quantum correlations beyond the mean-field level while preserving the model's simplicity. However, it also poses challenges including the loss of well-defined quantum numbers, which leads to problematic outcomes when computing observables other than the energy.  This is colloquially referred to as L\"owdin's dilemma, \cite{lykos1963discussion} which states that mean-field solutions can either adhere to good wavefunction or energy principles, but not both. 

One way to overcome L\"owdin's dilemma is by restoring the broken symmetries through projection.  This is often referred to as a Projection After Variation (PAV) approach, as it projects out the desired wave function components after the symmetry-breaking mean field optimization\cite{ring2004nuclear}.  Within the variational principle, this can be improved through the Variation After Projection approach (VAP), which exploits the fact that the optimal wavefunction at the mean-field level does not necessarily lead to the largest energetic gain after projection.  The spin-projected Extended Hartree--Fock (EHF)\cite{mayer:1980, mayer1980spin} and Projected Hartree-Fock approach\cite{scuseria2011projected,jimenez2012projected} are notable examples of the VAP on top of UHF in quantum chemistry, however the field of nuclear structure physics has seen considerably more developments in this direction \cite{launey2016symmetry, meng2006relativistic, bender2003self, zhao2016configuration, sheikh2021symmetry}.  

The key feature of an efficient VAP method is the existence of a computationally facile projector method.  Various approaches have been developed over the past decades, including the L\"owdin spin-filter\cite{lykos1963discussion} and Peierls--Yoccoz projection \cite{peierls1957collective} operator methods, among others \cite{cassamchenai:1998}.  Unfortunately, the exact implementation of these approaches involves effective $N$-body operations on the target mean-field state, reintroducing quantum correlations at a significantly steep computational cost.  Analytic investigations of the L\"owdin spin-filter operators has led to exact expansions of the projected wavefunctions in terms of Configuration State Functions (CSF) with tabulated Sanibel coefficients\cite{manne1966table, lowdin1955quantum}, however, due to the combinatorial nature of the CSF space, this approach has an unfavorable scaling for larger systems.  In contrast, the Peierls--Yoccoz projector takes a different stance within the Generator Coordinate Method (GCM) \cite{peierls1957collective,ring2004nuclear}, and projects on good quantum numbers by integrating over the degenerate Goldstone modes.  Unfortunately the overlap between any two degenerate states vanishes asymptotically with large system size, leading to issues of size-consistency\cite{jimenez2012projected}.  This unfavorable scaling instigated the development of approximate projection approaches.  For instance, shortly after the introduction of L\"owdin's spin filter projection operator, Amos  showed that filtering out a small subset of unwanted spin components already purifies the wavefunction to large extent.\cite{amos1964unrestricted} Another example is the Half-Projected Hartree-Fock (HPHF) method, which reduces the GCM to just a frugal linear combination of the original target and its spin-reversed counterpart, with remarkable accuracy. \cite{cox1976half, smeyers1978half, ruiz2022half}  The latter is obtained by simply substituting all spin-orbitals by their spin counterpart, effectively swapping all spin-up and spin-down orbitals.  

The main focus of the present paper is to revisit the process of spin-symmetry breaking and restoration in quantum chemistry by interfacing the spin-constrained Unrestricted Hartree-Fock (c-UHF) method with a Non-Orthogonal Configuration Interaction (NOCI) approach restricted to the configuration function space.  We will refer to this approach as the Spin-Projected constrained Unrestricted Hartree-Fock method (SPcUHF). The c-UHF method is a conventional UHF approach with the added feature that the expectation value of the spin is constrained to a certain preset value.\cite{andrews:1991,debaerdemacker:2023,ayati:2024}  This enables fine-tuned control of the spin symmetry breaking from the spin-zero Restricted Hartree-Fock (RHF) level to the UHF level in a controlled and continuous way.  Fitting within the framework of UHF approaches, the c-UHF has the benefit that each molecular orbital is a proper $SU(2)$ spinor.  This particular aspect paves the way to construct all possible spin-recoupled configuration functions from the original c-UHF state, and retain the spin-projected states either algebraically by means of explicit Racah spin recouplings, or computationally via a NOCI approach.  Because of the mutually non-zero orbital overlaps between the spin-up and spin-down sectors, we make an argument for the latter approach in the present paper.  We will discuss the SPcUHF as a VAP approach built upon a spin-symmetry broken UHF state, very much similar to the spin-projected EHF\cite{mayer:1980} and PHF.\cite{jimenez2012projected}  The main difference with the latter methods is that our NOCI allows for the automatic identification of the relevant SCFs, as well as their relative weights, without fixing them externally through the Sanibel coefficients.  Furthermore, by virtue of the controlled spin-symmetry breaking in the c-UHF formalism, we will show how a limited number of recoupled configurations is sufficient to effectively restore the spin symmetry and approximately capture the associated electron correlation energy.  As such, the present approach can also be regarded as a generalization of the HPHF method.


The paper is organized as follows.  In Section \ref{section:spinprojcuhf}, we succinctly recapitulate the key features of the c-UHF method, and introduce the spin-projection approach via a NOCI procedure.  By means of simple two- and four-electron examples, we provide intuition on the approximate projection in the restricted spin space.  In Section \ref{sec:results}, results for up to eight-electron systems are presented and discussed, and conclusions are formulated in Section \ref{sec:conclusions}.

\section{Spin-Projected c-UHF}\label{section:spinprojcuhf}

\subsection{Constrained-UHF}\label{subsection:cUHF}

The first step in the SPcUHF procedure is to produce a HF mean-field state in which the spin symmetry is broken in such a way that the proper spin-projection along the $z$-axis for each spin-orbital is maintained.  Consequently, the natural choice is the UHF formalism, in which both spin-up and spin-down sectors are not allowed to mix.  The key aspect of the spin-constrained c-UHF is to optimize the spin-orbitals in a single Slater determinant wavefunction subject to a user-defined expectation value of the spin. The spin constraint is introduced into the UHF formalism using a Lagrange multiplier ($\lambda$)
\begin{equation}
\hat{\mathcal{H}} = \hat{H} + \lambda [\hat{S}^{2} - S(S+1)].
\label{eq:Lag}
\end{equation}
\

By minimizing the expectation value of this constrained Hamiltonian, we aim to minimize the Coulomb energy $\expval{\hat{H}}$ while adhering to the spin constraint enforced by
\begin{equation}
\frac{\partial}{\partial \lambda} \mel{\textrm{c-UHF}(S)}{\hat{\mathcal{H}}}{\textrm{c-UHF}(S)} = \mel{\textrm{c-UHF}(S)}{\hat{S}^{2}}{\textrm{c-UHF}(S)} - S(S+1)\equiv0
\label{eq:constr}
\end{equation}
with $S(S+1)$ a user-imposed value, not necessarily corresponding to an integer valued $S$. 
Consequently, the optimized wavefunction $\ket{\textrm{c-UHF}(S)}$ and the associated expectation value of the physical Hamiltonian
\begin{equation}
\mathcal{E}(S) = \mel{\textrm{c-UHF}(S)}{\hat{H}}{\textrm{c-UHF}(S)}
\label{eq:lagexpener}
\end{equation}
are implicitly determined by the user-imposed value of the spin constraint $S$.

RHF and UHF emerge as special cases of c-UHF.  By imposing the spin constraint $S=0=S_{\textrm{RHF}}$, the c-UHF approach forces the Slater determinant to the RHF solution, whereas imposing the, initially unknown, $S=S_{\textrm{UHF}}$ spin will yield the UHF solution.  Since the UHF solution is the globally optimal Slater determinant over all possible expectation values of the spin operator, the global minimum of the $\mathcal{E}(S)$ function (eq.\ \ref{eq:lagexpener}) will also coincide with the UHF solution
\begin{equation}
    E_\text{UHF} = \min_S \mathcal{E}(S)
\end{equation}

Constrained Hartree-Fock methods have been investigated extensively in nuclear structure physics, \cite{bender2003self} and have been explored previously in quantum chemistry as well for different applications. \cite{mukherji:1963}

\subsection{Spin projection and recoupling}\label{subsection:recoupling}
\subsubsection{Spin tensorial properties}
The c-UHF Slater determinant of an $N$-electron system can be conveniently written as
\begin{equation}
    |\textrm{c-UHF}\rangle=\prod_{i=0}^{\frac{N}{2}-1}a_{2i\su}^\dag \prod_{j=0}^{\frac{N}{2}-1}a_{2j+1\sd}^\dag|\theta\rangle\label{projection:cuhf}
\end{equation} 
in which $a^\dag_{2i\su}$ and $a^\dag_{2j+1\sd}$ denote the creation of a spin-up ($\su$) and spin-down ($\sd$) electron in one of $2L$ spin-orbitals respectively, and $|\theta\rangle$ is the zero-particle vacuum.  For simplicity, we will assume that $N$ is even, and that the electrons are distributed equally over spin-up and spin-down spin-orbitals, however extensions of this framework can be considered.  For practical purposes that will become clear in later sections, we label the spatial components of the spin-up and spin-down orbitals by even ($2i$) and odd ($2j+1$) numbers, respectively.  All spatial orbitals are orthogonal within each individual spin sector, but they are not necessarily orthogonal between the two sectors.  This is reflected in the anticommutation relations
\begin{align}
     &\{a_{2i\su}^\dag,a_{2j\su}\}=\{a_{2i+1\sd}^\dag,a_{2j+1\sd}\}=\delta_{ij},\\
     &\{a^\dag_{2i\su},a_{2j+1\sd}\}=\{a^\dag_{2i\sd},a_{2j+1\su}\}=0,\\
     &\{a^\dag_{2i\su},a_{2j+1\su}\}=\{a^\dag_{2i\sd},a_{2j+1\sd}\}=G_{2i,2j+1},\quad\forall i,j=0\dots L-1\label{spin:tensor:updown}
\end{align}
 which complete the standard anticommutation relations
\begin{equation}
    \{a_{i\sigma}^\dag,a_{j\tau}^\dag\}=\{a_{i\sigma},a_{j\tau}\}=0,\quad\forall i,j=,0\dots 2L-1\quad\forall\sigma,\tau=\su,\sd.
\end{equation}
By virtue of the HF Roothaan procedure underneath the c-UHF state optimization, the spin-orbitals within each individual spin sector constitute a complete and orthonormal basis for the spatial component of the single-particle Hilbert space.  Consequently, the overlap matrix $\bm{G}$ is an orthogonal $L\times L$ matrix in general, reducing to the identity matrix $\bm{G}=\mathbb{I}$ 
only when the spin-up and spin-down orbitals are identical, up to a permutation of the orbital indices.  For convenience, we will order the spatial orbitals of each spin sector with increasing HF single-particle energy, such that the overlap matrix coincides identically with $\bm{G}\equiv\mathbb{I}$ for an RHF state.  Also, rows and columns will explicitly be denoted by even/odd indices $2i$, $2j+1$ respectively, referring to the spin $\uparrow$ and $\downarrow$ sectors of UHF.  For instance, the overlap between the lowest spin-up and spin-down orbital will be $G_{01}$, and diagonal matrix elements of $\bm{G}$ will be denoted by $G_{2i,2i+1}$.

The total spin operators are covariant with respect to unitary transformations between spatial orbital, so they can be expressed in either the spin-up or spin-down, or an unrelated and unbiased third orthonormal basis, defined by the anticommutation relations
\begin{equation}
    \{b_{i\sigma}^\dag,b_{j\tau}^\dag\}=\{b_{i\sigma},b_{j\tau}\}=0,\quad\{b_{i\sigma}^\dag,b_{j\tau}\}=\delta_{ij}\delta_{\sigma\tau}.\quad\forall ij=1\dots L,\quad\&\quad \sigma,\tau=\su,\sd.\label{spin:banticommutator}
\end{equation}
It is always possible to express the c-UHF single-particle operators in this third basis via a unitary transformation
\begin{equation}
    a^\dag_{2i\su}=\sum_{j=1}^L U^\su_{ij}b^\dag_{j\su},\quad a^\dag_{2i+1\sd}=\sum_{j=1}^L U^\sd_{ij}b^\dag_{j\sd},\label{spin:aUb}
\end{equation}
with the unitary matrices $\bm{U}^\su$ and $\bm{U}^\sd$ related via the overlap matrix 
\begin{equation}
    \bm{U}^\su (\bm{U}^\sd)^\dag = \bm{G}.
\end{equation}
The spin operators in this third orthonormal basis become
\begin{equation}
    \hat{S}^+ = \sum_{i=1}^L b_{i\su}^\dag b_{i\sd},\quad \hat{S}^- = (\hat{S}^+)^\dag  = \sum_{i=1}^L b_{i\sd}^\dag b_{i\su},\quad \hat{S}_0=\tfrac{1}{2}\sum_{i=1}^L[b_{i\su}^\dag b_{i\su} - b_{i\sd}^\dag b_{i\sd}],\label{spin:definition}
\end{equation}
spanning an $SU(2)$ algebra 
\begin{equation}
    [\hat{S}_0,\hat{S}^\pm]=\pm\hat{S}^\pm,\quad [\hat{S}^+,\hat{S}^-]=2\hat{S}_0,
\end{equation}
with the total spin operator as the Casimir operator
\begin{equation}
    \hat{\vec{S}}^2=\hat{S}_0^2 + \tfrac{1}{2}[\hat{S}^+ \hat{S}^- + \hat{S}^-\hat{S}^+].\label{spin:casimir}
\end{equation}
With the relations (\ref{spin:aUb}) and definitions (\ref{spin:banticommutator}) \& (\ref{spin:definition}), it is straightforward to verify the $SU(2)$ spinor properties \cite{talmi:1993} of the c-UHF spin-orbitals for all even ($\su$) and odd ($\sd$) indices $i$ 
\begin{align}
    [\hat{S}^\pm,a^\dag_{im}]&=\sqrt{(\tfrac{1}{2}\mp m)(\tfrac{1}{2}\pm m+1)}a^\dag_{im\pm 1}\\
    [\hat{S}_0,a^\dag_{im}]&=ma^\dag_{im}
\end{align}
in which the notation $m=+\frac{1}{2}$ and $m=-\frac{1}{2}$ has been used to denote the spin-up and spin-down sector respectively.  As a result, the spin algebra eq.\ (\ref{spin:definition}) allows us to flip the spin of any c-UHF spin-orbital.  Moreover, they transform as genuine spinors under $SU(2)$ rotations\cite{talmi:1993}
\begin{equation}
    \hat{R}(\Omega)a^\dag_{im} \hat{R}(\Omega)^{-1}=\sum_{m^\prime=\pm\frac{1}{2}}a^\dag_{im^\prime}\mathcal{D}^{\frac{1}{2}}_{m^\prime m}(\Omega)\label{spin:proj:spinor}
\end{equation}
in which $\hat{R}(\Omega)$ is an $SU(2)$ rotation over the Euler angles $\Omega=\{\alpha,\beta,\gamma\}$ and $\mathcal{D}^{S}_{m^\prime m}$ are the Wigner-$\mathcal{D}$ $SU(2)$ irreducible representations of dimension $2S+1$ and projection $(m^\prime,m)$ in the laboratory and intrinsic reference frame $z$-axis respectively. \cite{rose:1957}

\subsubsection{Spin projection}
The c-UHF state (\ref{projection:cuhf}) is not an eigenstate of the $SU(2)$ spin operator (\ref{spin:casimir}), so it needs to be projected on a state with proper quantum numbers.  This can be achieved using L\"owdin's filter operator
\begin{equation}
    \hat{P}^S=\prod_{S^\prime\neq S}\tfrac{\hat{S}^2-S^\prime(S^\prime+1)}{S(S+1)-S^\prime(S^\prime+1)},\label{spin:proj:lowdin}
\end{equation}
in which each spin component is filtered out except for the desired $S$.  From the definition of the spin-operator eq.\ (\ref{spin:casimir}), it is immediately clear that the L\"owdin operator eq.\ (\ref{spin:proj:lowdin}) is a $2(N-1)$-body operator with hitherto unfavourable computational scaling. 

A different approach is provided by the Peierls--Yoccoz projector\cite{peierls1957collective}, which projects out the state on the desired irreducible representation by integrating over all possible $SU(2)$ group actions $\hat{R}(\Omega)$\cite{ring2004nuclear}
\begin{equation}
    \hat{P}^S_{MK}=\int {\mathcal{D}^{S}_{MK}}(\Omega)^\ast\hat{R}(\Omega) d\Omega. \label{spin:proj:gcm}
\end{equation}
In order to better appreciate the spin projection properties of the operator (\ref{spin:proj:gcm}) while also setting the stage for the SPcUHF, it is instructive to consider its action on a two-electron c-UHF state, e.g.\ for the \ce{H_2} molecule
\begin{equation}
    \hat{P}^S_{MK}a_{0,\frac{1}{2}}^\dag a_{1,-\frac{1}{2}}^\dag|\theta\rangle=\int d\Omega \mathcal{D}^{S}_{MK}(\Omega)^\ast\hat{R}(\Omega)a_{0,\frac{1}{2}}^\dag\hat{R}(\Omega)^{-1}\hat{R}(\Omega) a_{1,-\frac{1}{2}}^\dag\hat{R}(\Omega)^{-1}\hat{R}(\Omega)|\theta\rangle
\end{equation}
in which we have twice introduced the resolution of the identity $\hat{R}(\Omega)^{-1}\hat{R}(\Omega)=\mathbb{I}$.  Now we can exploit the $s=\frac{1}{2}$ spinor properties (\ref{spin:proj:spinor}) of the spin-orbitals, as well as the scalar properties of the particle vacuum to arrive at
\begin{equation}
    \hat{P}^S_{MK}a_{0,\frac{1}{2}}^\dag a_{1,-\frac{1}{2}}^\dag|\theta\rangle=\sum_{m_0=\pm\frac{1}{2}}\sum_{m_1=\pm\frac{1}{2}}a^\dag_{0,m_0}a^\dag_{1,m_1}|\theta\rangle\int d\Omega  \mathcal{D}^{S}_{MK}(\Omega)^\ast \mathcal{D}^{\frac{1}{2}}_{m_0,\frac{1}{2}}(\Omega)\mathcal{D}^{\frac{1}{2}}_{m_1,-\frac{1}{2}}(\Omega).
\end{equation}
The reduction and orthogonality relations of the Wigner-$\mathcal{D}$ matrices\cite{rose:1957} can now be used to give
\begin{align}
    &\mathcal{D}^{j_1}_{m_1k_1}(\Omega)\mathcal{D}^{j_2}_{m_2k_2}(\Omega)=\sum_{j,mk}\langle j_1m_1,j_2m_2|jm\rangle\langle j_1k_1,j_2k_2|jk\rangle \mathcal{D}^{j}_{mk}(\Omega),\\
    &\int d\Omega \mathcal{D}^{j_1}_{m_1k_1}(\Omega)^\ast\mathcal{D}^{j_2}_{m_2k_2}(\Omega) = \frac{(4\pi)^2}{2(2j_1+1)}\delta_{j_1j_2}\delta_{m_1m_2}\delta_{k_1k_2},
\end{align}
with $\langle j_1m_1,j_2m_2|jm\rangle$ denoting the conventional Clebsch--Gordan (CG) coefficients. \cite{rose:1957}  We obtain the final result
\begin{equation}
     \hat{P}^S_{MK}a_{0,\frac{1}{2}}^\dag a_{1,-\frac{1}{2}}^\dag|\theta\rangle=\tfrac{(4\pi)^2}{2(2S+1)}\langle\tfrac{1}{2}\tfrac{1}{2},\tfrac{1}{2}-\tfrac{1}{2}|SK\rangle [a_0^\dag a_1^\dag]^{S}_M|\theta\rangle,\label{spin:proj:twoelec}
\end{equation}
which indeed clarifies that the projection operator produces a spin-recoupled state 
\begin{equation}
   [a_0^\dag a_1^\dag]^{S}_M|\theta\rangle=\sum_{m_0,m_1=\pm\frac{1}{2}}\langle \tfrac{1}{2}m_0,\tfrac{1}{2}m_1|SM\rangle a^\dag_{0m_0}a^\dag_{1m_1}|\theta\rangle, \label{spin:proj:2ecg}
\end{equation}
with the desired spin $S$ and projection $M$.  For our two-electron system, only the $S=0$ singlet and $S=1$ triplet are allowed.  Focusing on the $M=0$ sector only, we see that both the original c-UHF state $a^\dag_{0,+\frac{1}{2}}a^\dag_{1,-\frac{1}{2}}|\theta\rangle=a^\dag_{0\su}a^\dag_{1\sd}|\theta\rangle$ and its symmetry-broken partner $a^\dag_{0,-\frac{1}{2}}a^\dag_{1,+\frac{1}{2}}|\theta\rangle=a^\dag_{0\sd}a^\dag_{1\su}|\theta\rangle$ are involved in the projected state.  This is remarkable as the latter had not been explicitly generated at the c-UHF optimization stage.  The observation that both configurations are needed to reconstruct the spin-symmetric two-electron state forms the basis of the HPHF method. \cite{sheikh2021symmetry, cox1976half, smeyers1978half}  In HPHF, the spin-symmetric state is approximated by a 50/50 equally weighted linear combination of the UHF state and its energetically degenerate symmetry-broken partner.  Our back-on-the-envelope analysis for the two-electron case shows that these weight factors can be explained as CG coefficients in the projected state (\ref{spin:proj:2ecg}).  

Moving towards multiple electron systems, it is interesting to explore the potential of direct recouplings with CG coefficients to construct exact symmetry-projected states from c-UHF seed states.  To demonstrate this idea, the four-electron c-UHF (\ref{projection:cuhf}) can be reordered as 
\begin{equation}
    a_{0\su}^\dag a_{1\sd}^\dag a^\dag_{2\su}a^\dag_{3\sd}|\theta\rangle
\end{equation}
up to an overall minus sign phase factor.  The action of the projection operator (\ref{spin:proj:gcm}) on this state yields
\begin{equation}
    \hat{P}_{MK}^S a_{0\su}^\dag a_{1\sd}^\dag a^\dag_{2\su}a^\dag_{3\sd}|\theta\rangle=\tfrac{(4\pi)^2}{4(2S+1)}\delta_{K0}\sum_{S_{01},S_{23}}(-)^{S_{01}+S_{23}}\langle S_{01} 0,S_{23},0|S0\rangle [[a_0^\dag a_1^\dag]^{S_{01}}[a_2^\dag a_3^\dag]^{S_{23}}]^{S}_M|\theta\rangle.
\end{equation}\label{spin:proj:4electrons}
The interpretation of this state is that electrons 0 and 1 are first recoupled to an intermediate singlet or triplet spin $S_{01}$, electrons 2 and 3 are recoupled to an intermediate singlet or triplet spin $S_{23}$, which are subsequently recoupled to the total spin $S$. The allowed total spin values can range over $S=0,1,2$, with different multiplicities $(2,3,1)$.  At this point, it appears as if spin projection is a solved algebraic problem in terms of tabulated CG\cite{rose:1957} or Sanibel coefficients\cite{manne1966table}.  However, an important caveat is that the spin-orbitals of the two spin-sectors are not mutually orthogonal (\ref{spin:tensor:updown}), which will affect the weighting of the recoupled electron pairs. Again, this is best understood in the two-electron case.  Consider the $S=1$ projected two-electron state (\ref{spin:proj:twoelec}) with $(M,K)=(0,0)$
\begin{equation}
    \hat{P}^1_{00}a_{0\su}^\dag a_{1,\sd}^\dag|\theta\rangle=\tfrac{(4\pi)^2}{6\sqrt{2}} [a_{0}^\dag a_{1}^\dag]^1_0|\theta\rangle,
\end{equation}
Since each spin-sector constitutes a complete basis set for the spatial part of the orbitals, it is always possible to express $a^\dag_{1\sd}$ in the (spin-flipped) spin-up basis
\begin{equation}
    a^\dag_{1\sd} = \sum_{i=0}^{L-1}G_{2i,1}a^\dag_{2i\sd},
\end{equation}
with $\bm{G}$ the overlap defined in (\ref{spin:tensor:updown}).  The $S=1$ projection then becomes
\begin{equation}
       \hat{P}^1_{00}a_{0\su}^\dag a_{1,\sd}^\dag|\theta\rangle=\tfrac{(4\pi)^2}{6\sqrt{2}}\sum_{i=0}^{L-1} G_{2i,1}[a_{0}^\dag a_{2i}^\dag]^1_0|\theta\rangle=\tfrac{(4\pi)^2}{6\sqrt{2}}\sum_{i=1}^{L-1} G_{2i,1}[a_{0}^\dag a_{2i}^\dag]^1_0|\theta\rangle,
\end{equation}
in which the $i=0$ term vanishes identically in the last equality as it is impossible to create a $S=1$ triplet state with the same spatial orbitals.  In other words, a portion proportional to the overlap ${G}_{01}$ disappears from the $S=1$ projected state, due to the Pauli exclusion principle.  Moreover, in the extreme case of RHF in which spin-down and spin-up spatial orbital are the same (${G}_{01}=1)$, the projection operator completely annihilates the RHF state because there is no $S=1$ component present in the RHF wavefunction
\begin{equation}
       \hat{P}^1_{00}a_{0\su}^\dag a_{0,\sd}^\dag|\theta\rangle=0.
\end{equation}
The existence of these vanishing components complicates the algebraic formulation of the projection operator in terms of CG recoupling coeficients.  Eventually, the relative weights of the SCFs resulting from the  projection should be flexible enough, for instance for size consistency purposes in bond-dissociation purposes.  Therefore, an alternative projection method is desirable to facilitate this flexibility.
\subsection{NOCI}
An alternative approach to generate the spin-projected states in the presence of non-orthogonal spin-orbitals is to retain all possible configurations that would occur in a CG recoupling scheme and use these to build a basis for Non-Orthogonal Configuration Interaction (NOCI). \cite{thom2009hartree, malmqvist1986calculation,sundstrom2014,burton2019}  We will refer to NOCI computations as NOCI($k$) in which $k$ denotes the total number of configurations.   For a four-electron system, there are six possible ways in which spin-up and spin-down electrons can be distributed over four spatial orbitals, leading to a NOCI(6) subspace containing the configurations
\begin{align}
    \{& 
    a_{0\su}^\dag a_{1\sd}^\dag a^\dag_{2\su}a^\dag_{3\sd}|\theta\rangle,\ 
    a_{0\su}^\dag a_{1\su}^\dag a^\dag_{2\sd}a^\dag_{3\sd}|\theta\rangle,\
    a_{0\sd}^\dag a_{1\su}^\dag a^\dag_{2\su}a^\dag_{3\sd}|\theta\rangle,\notag\\
    &a_{0\su}^\dag a_{1\sd}^\dag a^\dag_{2\sd}a^\dag_{3\su}|\theta\rangle,\
    a_{0\sd}^\dag a_{1\sd}^\dag a^\dag_{2\su}a^\dag_{3\su}|\theta\rangle,\
    a_{0\sd}^\dag a_{1\su}^\dag a^\dag_{2\sd}a^\dag_{3\su}|\theta\rangle\} . \label{spin:noci:6states}
\end{align}
Because all spin-orbitals have good spinor properties, they are guaranteed to retain good spin-tensorial properties in the recouplings, however the weightings of the recoupled terms will depend on the overlaps between the spin-up and spin-down sectors.  It is shown explicitly in Appendix \ref{section:appendix} that the NOCI states for four electrons are compatible with the CG recouplings in the presence of non-orthogonal overlaps.

In cases where there are spin-up and spin-down electrons occupying the same spatial orbitals in \(|\textrm{c-UHF}(S) \rangle\), some of the related configurations will end up with identical spin-orbitals residing in a single spin sector, causing some determinants in the NOCI wavefunction to vanish. For example, in the four-electron case, if spin orbitals 0 and 1 are the same or ``paired'' at the UHF level between spin-up and spin-down sectors, two of the projected spin configurations \(a_{0\su}^\dag a_{1\su}^\dag a_{2\sd}^\dag a_{3\sd}^\dag |\theta\rangle = a_{0\su}^\dag a_{0\su}^\dag a_{2\sd}^\dag a_{3\sd}^\dag |\theta\rangle = 0\) and \(a_{0\sd}^\dag a_{1\sd}^\dag a_{2\su}^\dag a_{3\su}^\dag |\theta\rangle = a_{0\sd}^\dag a_{0\sd}^\dag a_{2\sd}^\dag a_{3\sd}^\dag |\theta\rangle = 0\) vanish because the same spin-orbitals (0 and 1) are assigned to the same spin sector. Additionally, determinants like \(a_{0\su}^\dag a_{1\sd}^\dag a_{2\su}^\dag a_{3\sd}^\dag |\theta\rangle = a_{0\su}^\dag a_{0\sd}^\dag a_{2\su}^\dag a_{3\sd}^\dag |\theta\rangle\) and \(a_{0\sd}^\dag a_{1\su}^\dag a_{2\su}^\dag a_{3\sd}^\dag |\theta\rangle = a_{0\sd}^\dag a_{0\su}^\dag a_{2\su}^\dag a_{3\sd}^\dag |\theta\rangle\) are equivalent, as swapping degenerate spin-orbitals leaves the determinant unchanged. The same applies to \(a_{0\sd}^\dag a_{1\su}^\dag a_{2\sd}^\dag a_{3\su}^\dag |\theta\rangle\) and \(a_{0\su}^\dag a_{1\sd}^\dag a_{2\sd}^\dag a_{3\su}^\dag |\theta\rangle\). As a result, the NOCI(6) calculation reduces to NOCI(2), retaining only the determinants \(a_{0\su}^\dag a_{1\sd}^\dag a_{2\su}^\dag a_{3\sd}^\dag |\theta\rangle\) and \(a_{0\su}^\dag a_{1\sd}^\dag a_{2\sd}^\dag a_{3\su}^\dag |\theta\rangle\). In general, for a given symmetry-broken c-UHF state \(|\textrm{c-UHF}(S)\rangle\), the effective number $k_{\text{eff}}$ of projected configurations remaining in the NOCI($k_{\text{eff}}$) basis depends on the number of unpaired spin-orbitals (\(M\)) across the spin sectors, following the binomial coefficient \(k_{\text{eff}} = \binom{M}{M/2}\). The procedure is outlined in algorithm \ref{algo:pairing}.  The number of unpaired spatial orbitals $M$ can be inferred directly from the number of non-1 diagonal elements of $\bm{G}$, as $k_{\text{eff}}$ can be directly determined by the orthogonality of the NOCI basis which is a required step of a practical NOCI implementation.

\begin{algorithm}[H]
\caption{NOCI basis}\label{algo:pairing}
\begin{algorithmic}[1]
    \Require{$k$,$|\textrm{c-UHF}(S) \rangle$}\Comment{}
    \State $\textrm{NOCIBasis}(k_{\text{eff}}) \gets$ \Call{GenNOCIBasis}{$k,|\textrm{cUHF}\rangle$}\Comment{$k_{\text{eff}} \gets \binom{M}{M/2}$} 
    \State $E\gets$ \Call{NOCI}{$k_{\text{eff}}$, NOCIBasis($k_{\text{eff}}$)}
\end{algorithmic}
\end{algorithm}

We also need to be mindful of the nonorthogonality of orbitals involved in the spin flip. While a full spin flip ensures orthogonality within the spin-up and spin-down orbital sets, our protocol does not automatically guarantee this. To perform NOCI correctly, and potentially PT2 perturbative corrections, it's essential first to orthogonalize the orbitals within each spin sector for all the projected configurations.

We are now in the position to discuss the algorithmic steps in the spin-projected c-UHF method (SPcUHF).  The goal of the SPcUHF is to compute the ground-state energy of a molecular system through the variational principle.  The initial step is to identify the possible range of spin expectation values $\langle S^2\rangle$ for a c-UHF state, which only depends on the number of electrons $N$  \cite{cassam2015spin}

\begin{equation}
    0 \leq \langle \textrm{c-UHF}| S^2|\textrm{c-UHF}\rangle \leq \frac{N}{2}.
\end{equation}
We then apply a global variational optimization.  For each constrained $\langle\textrm{c-UHF}|\hat{S}^2|\textrm{c-UHF}\rangle$ we 
compute a c-UHF ground-state energy $\mathcal{E}_{\textrm{c-UHF}}(S)$ 
using eq.~\eqref{eq:lagexpener}, and project the associated $|\textrm{c-UHF}(S)\rangle$ state onto proper spin eigenstates using the NOCI approach outlined above.  See Algorithm \ref{algo:spcuhf} for an outline in pseudocode.
The resulting NOCI energy $E_{\textrm{NOCI}}$ will be lower (but not strictly) than the reference c-UHF energy $E_{\textrm{c-UHF}}(S)$. Finally, we retain the lowest $E_{\textrm{NOCI}}$ over all expectation values $\langle\textrm{c-UHF}|S^2|\textrm{c-UHF}\rangle$ as the variational minimum.  In all figures below, we will report energy profiles of both the c-UHF and SPcUHF/NOCI methods as a function of the imposed spin constraint $\langle\textrm{c-UHF}|S^2|\textrm{c-UHF}\rangle$ that went into the seed c-UHF state, although all resulting SPcUHF states are technically free from spin contamination by construction.

\begin{algorithm}[H]
\caption{spin-projected c-UHF}\label{algo:spcuhf}
\begin{algorithmic}[1]
    \Require{$N$,$\langle S^2\rangle = \langle\textrm{c-UHF}|S^2|\textrm{c-UHF}\rangle$}\Comment{Number of particles in the system and imposed spin expectation value}
    \Ensure{$E$}\Comment{selected energy eigenvalues}
    \Procedure{spin-projected c-UHF}{}
    \State $|\textrm{c-UHF}\rangle \gets$ \Call{cUHF}{$N$,$\langle S^2\rangle$}
    \State $k\gets \binom{N}{N/2}$ \label{algo:k}
    \State $\textrm{NOCIBasis}(k_{\textrm{eff}}) \gets$ \Call{GenNOCIBasis}{$k,|\textrm{cUHF}\rangle$}\Comment{construct and orthogonalize NOCI basis} 
    \State $E\gets$ \Call{NOCI}{$k_{\textrm{eff}}$, NOCIBasis($k_{\textrm{eff}}$)}
    \EndProcedure
\end{algorithmic}
\end{algorithm}

\begin{figure}[htb]
\centering
\begin{subfigure}{.5\textwidth}
  \centering
  \includegraphics[scale=1.0]{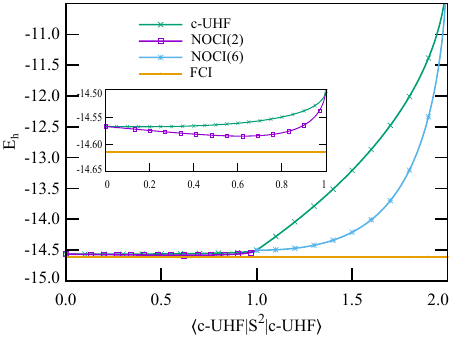}
  \caption{}
  \label{fig:Be_631g_NOCI}
\end{subfigure}%
\begin{subfigure}{.5\textwidth}
  \centering
  \includegraphics[scale=1.0]{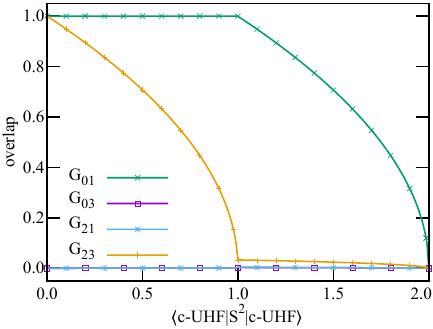}
  \caption{}
  \label{fig:Be_631g_overlap}
\end{subfigure}
\caption{For Be/6-31G: (a) Comparison of c-UHF, NOCI(2), and NOCI(6) energies with FCI energy across the full range of imposed constraint $\langle S^2\rangle$. (b) The overlap between $\alpha$ (0, 2) and $\beta$ (1, 3) molecular spin-orbitals.}
\label{fig:Be_631g}
\end{figure}
An illustrative four-electron example is provided by the \ce{Be} atom in 6-31G basis set, presented in Figure~\ref{fig:Be_631g}.  Before discussing this in more detail, it is worth recalling that the quantum correlations in \ce{Be} are predominantly of dynamical nature, so one should not expect results of quantitative nature from spin-symmetry breaking approaches which are mainly tailored for systems dominated by static correlation, such as the Paldus \ce{H_4} system discussed in Section~\ref{subsection:4e}.  Nevertheless, \ce{Be}/6-31G serves as an excellent example to discuss the general features of the SPcUHF method.  Figure~\ref{fig:Be_631g_NOCI} shows the c-UHF energy and NOCI energy as a function of the imposed $\langle\textrm{c-UHF}|S^2|\textrm{c-UHF}\rangle$ value.  The NOCI results of interest are labelled as NOCI($k$), with $k$ referring to the number of basis states (\ref{spin:noci:6states}) that went into the diagonalization procedure. Since there are four electrons in the system, we expect NOCI(6) to fully restore the spin symmetry.  

The first aspect to notice is that the c-UHF curve has a global minimum at $\langle\textrm{c-UHF}|S^2|\textrm{c-UHF}\rangle=0$, confirming that symmetry breaking does not occur spontaneously in UHF theory. In the range $\langle\textrm{c-UHF}|S^2|\textrm{c-UHF}\rangle\in[0,1]$, the c-UHF keeps the 1s $\alpha$ and $\beta$ spin-orbitals paired, demonstrated by $G_{01} = 1$ in Figure~\ref{fig:Be_631g_overlap}, and thus the NOCI(6) results reduce to NOCI(2) in the first interval.  Notably, the SPcUHF results show a modest variational minimum at a non-zero $\langle\textrm{c-UHF}|S^2|\textrm{c-UHF}\rangle=0.62$, capturing 37.69\% of the correlation energy that could not be accounted for at the UHF level. This result illustrates that, although \ce{Be} is dominated by dynamical correlations, there is a non-negligible portion of static correlation present as well, which could be captured through the tunable symmetry breaking obtained through the c-UHF approach.

\subsection{Projection-space restricted NOCI} \label{sec:PSrNOCI}

The NOCI reformulation of the spin projection operator (\ref{spin:proj:gcm}) provides an exact algebraic finite-size reformulation of a continuous generator coordinate method problem.  However in the most general formulation, the present approach suffers from a pernicious computational scaling, due to the size of the generated NOCI basis, as seen in line \ref{algo:k} of Algorithm \ref{algo:spcuhf}.  For a system of $N$ electrons, there are $k=\binom{N}{N/2}$ ways to generate all possible spin configurations that will occur in the CG expansion.  Although this scaling is independent of the size of the basis set $L$, and slightly better than full Configuration Interaction (FCI), it does not present a viable option for a many-body approach.  However, closer inspection of the derivative discontinuity in the NOCI curves around $\langle\textrm{c-UHF}|S^2|\textrm{c-UHF}\rangle=1$ in Figure \ref{fig:Be_631g_NOCI} reveals a potential physical mechanism to mitigate the exponential scaling based on the process of gradual symmetry breaking.  This process can be best understood starting at the $\langle\textrm{c-UHF}|S^2|\textrm{c-UHF}\rangle=0$ solution, corresponding to the RHF state.  At this point, there is no symmetry breaking and the spin-up and spin-down spatial orbitals are identical.  This is confirmed in Figure \ref{fig:Be_631g_overlap}, where the overlap matrix elements ${G}_{2i,2j+1}$ in the c-UHF method are presented as a function of imposed $\langle\textrm{c-UHF}|S^2|\textrm{c-UHF}\rangle$ valune.  At the $\langle\textrm{c-UHF}|S^2|\textrm{c-UHF}\rangle=0$ point,  all the spatial orbitals in the two spin sectors are paired, giving ${G}_{01}={G}_{23}=1$ as expected.  

\begin{figure}[b]
\centering
  \includegraphics{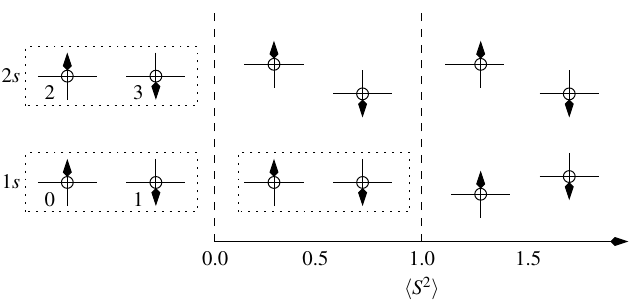}
  \caption{Interpretation of the pair-wise symmetry-breaking process in Be atom. The spin-up and spin-down spin-orbitals are labeled by even and odd numbers, respectively.  Dotted rectangles around pairs of spin-orbitals indicate that they are identical or paired.}\label{fig:bepairbreaking}
\end{figure}

An interesting phenomenon now occurs as we impose a non-zero $\langle\textrm{c-UHF}|S^2|\textrm{c-UHF}\rangle$ value.  It is clear from the behavior of ${G}_{01}$ and ${G}_{23}$ that the symmetry breaking is concentrated in the spin-orbitals 2 and 3, which are associated with the $2s$ valence electrons of \ce{Be}, whereas the $1s$ core orbitals remain paired, hence the appearance of NOCI(2) in this region.  It is only after the valence pair of electrons is maximally broken at $\langle\textrm{c-UHF}|S^2|\textrm{c-UHF}\rangle=1$ that the core electrons need to participate in the symmetry breaking as well.  The physics of this process can be well understood as it takes less energy to break a pair of electrons from the valence shell compared to the core.  Moreover, the majority of static correlation, however little present in \ce{Be}, should be found in the valence pair of electrons, explaining why the global minimum is found in the $\langle\textrm{c-UHF}|S^2|\textrm{c-UHF}\rangle\in[0,1]$ interval.  A cartoon interpretation of the pair-wise symmetry-breaking process is provided in Figure \ref{fig:bepairbreaking}.

The present observation opens up opportunities to reduce the computational scaling of the SPcUHF.  Since it is now understood that the critical spin-breaking happens only in the valence space, Figure \ref{fig:Be_631g} shows that a NOCI(2), in which only the two configurations in the CG expansion of the electron pair in the valence shell are considered is enough to restore the broken symmetry and provide the variational minimum.  This observation can serve as inspiration for an alternative Algorithm \ref{algo:spacerestrictedspcuhf} with a significantly reduced computational scaling.  In this algorithm, for each $\langle\textrm{c-UHF}|S^2|\textrm{c-UHF}\rangle$ interval starting from $\langle\textrm{c-UHF}|S^2|\textrm{c-UHF}\rangle\in[0,1]$, we restore the spin symmetry using NOCI($k_{\textrm{eff}}$) with $k_{\textrm{eff}}=\binom{M}{M/2}$ where $M$ is the number of unpaired spin-orbitals between the two spin sectors until either no gain in correlation energy is observed by going to the next interval, or the worst case $\langle\textrm{c-UHF}|S^2|\textrm{c-UHF}\rangle\in [\frac{N}{2}-1,\frac{N}{2}]$ is reached. It is to be expected that most of static correlation will be captured as soon as the valence electrons are exhausted.

\begin{algorithm}[H]
\caption{projection-space restricted spin-projected c-UHF}\label{algo:spacerestrictedspcuhf}
\begin{algorithmic}[1]
    \Require{$N$}\Comment{Number of particles in the system}
    \Ensure{$E$}\Comment{selected energy eigenvalues}
    \State $\textrm{ranges} = \{[0,1], [1,2], ..., [\frac{N}{2}-1,\frac{N}{2}]\}$  
    \Procedure{projection-space restricted spin-projected c-UHF}{}
    \While{[$\Delta E<0\wedge \textrm{range} \in \textrm{ranges}$]} \Comment{Increase space while correlation energy is being gained}
    \State $\langle S^2\rangle\in \textrm{range}$
    \State $|\textrm{c-UHF}\rangle\gets$ \Call{cUHF}{$N$,$\langle S^2\rangle$} \Comment{Precompute all c-UHF states}
    \State $k\gets \binom{n}{n/2}$\Comment{$k=2,6,20,70, 252, 924, 3432, 12870\dots$}
    \State $\textrm{NOCIBasis}(k_{\text{eff}}) \gets$ \Call{GenNOCIBasis}{$k,|\textrm{cUHF}\rangle$}\Comment{$k_{\text{eff}} \gets \binom{M}{M/2}$}
    \State $E\gets \min_{\langle S^2\rangle}$ \{\Call{NOCI}{$k_{\text{eff}}$,$\textrm{NOCIBasis}(k_{\text{eff}})$}\}
     \EndWhile
    \EndProcedure
\end{algorithmic}
\end{algorithm}

\subsection{Computational Details}\label{subsection:compdetail}

All c-UHF computations were performed using an in-house Python implementation, \cite{qunbrepo} and the NOCI were done with a customized Q-Chem package\cite{shao2015advances} with the LibGNME library for computing the necessary nonorthogonal matrix elements. \cite{BurtonLibGNME1,BurtonLibGNME2,libgnme}
In practice, the SPcUHF approach mainly introduces static electron correlation. 
Additional dynamic correlation can be incorporated as a second-order perturbative correction using the NOCI-PT2 approach.\cite{Burton2020}
This method defines a rigorous nonorthogonal extension to conventional 
multireference perturbation theories, and can provide quantitative accuracy
for molecular binding energies.
Prior to computing the NOCI-PT2 correction, pseudo-canonical occupied and virtual orbitals are constructed for 
each determinant in the NOCI expansion by diagonalizing the occupied-occupied and virtual-virtual blocks of the corresponding 
Fock matrix for that determinant.

\section{Results and Discussion}\label{sec:results}

We will now further explore the SPcUHF in full and reduced projection spaces for several four-, six-, and eight-electron systems that are archetypal for specific types of electron correlation.

\subsection{4-electron systems: \ce{LiH}/6-31G and \ce{H_4}/6-31G}\label{subsection:4e}

\begin{figure}[htb]
\centering
\begin{subfigure}{0.5\textwidth}
  \centering
  \includegraphics[scale=1.0]{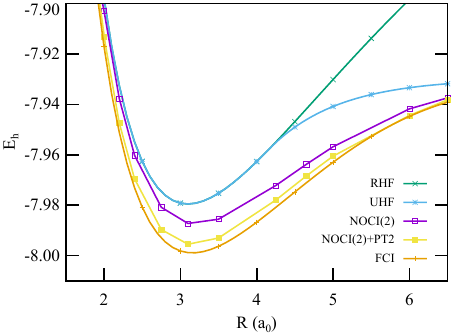}
  \caption{}
  \label{fig:LiH_ RUHFpes}
\end{subfigure}%
\begin{subfigure}{0.5\textwidth}
  \centering
  \includegraphics[scale=1.0]{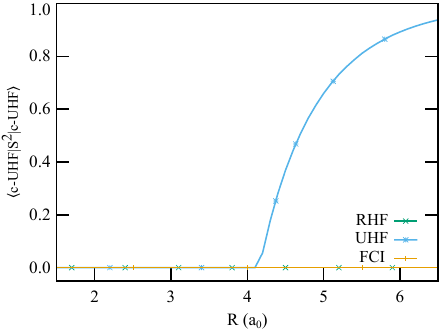}
  \caption{}
  \label{fig:LiH_RUHFspin}
\end{subfigure}
\caption{Total energy (a) and $\langle S^2 \rangle$ (b) of the RHF, UHF, SPcUHF, and FCI  ground states for the \ce{LiH}/6-31G binding curve.}
\label{fig:LiH_631g_ruhf}
\end{figure}

As a single-bond breaking process, the binding curve of \ce{LiH} in the 6-31G basis set strongly resembles the \ce{H_2} binding curve, but with a total of four electrons instead of two.  
Figure~\ref{fig:LiH_631g_ruhf} compares the ground-state potential energy surfaces for RHF, UHF, SPcUHF, and FCI. In the equilibrium regime, UHF gives the same result as RHF since the spin symmetry is not spontaneously broken at these points. However, as the bond length extends beyond the Coulson-Fischer point at $R=4.2a_{0}$, UHF consistently provides lower energies than RHF. Figure~\ref{fig:LiH_RUHFspin} plots the expectation value $\langle S^2 \rangle$ for the RHF, UHF, and FCI ground state along the bond dissociation, highlighting the symmetry-breaking present in UHF, with $\langle\textrm{UHF}|S^2|\textrm{UHF}\rangle$ becoming non-zero beyond $4.2\,a_{0}$. 

\begin{figure}[b]
\centering
\begin{subfigure}{.5\textwidth}
  \centering
  \includegraphics[scale=1.0]{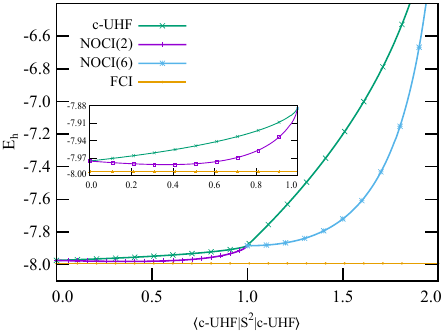}
  \caption{}
  \label{fig:LiH_275_NOCI}
\end{subfigure}%
\begin{subfigure}{.5\textwidth}
  \centering
  \includegraphics[scale=1.0]{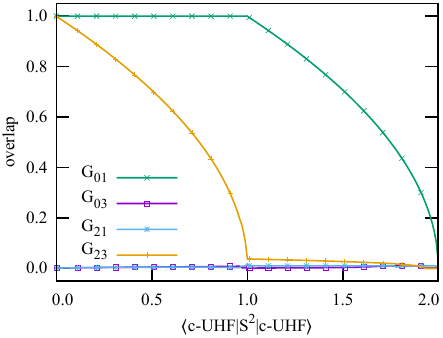}
  \caption{} 
  \label{fig:LiH_275_MOoverlap}
\end{subfigure}
\begin{subfigure}{.5\textwidth}
  \centering
  \includegraphics[scale=1.0]{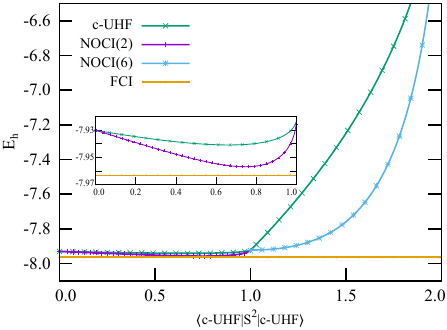}
  \caption{} 
  \label{fig:LiH_500_NOCI}
\end{subfigure}%
\begin{subfigure}{.5\textwidth}
  \centering
  \includegraphics[scale=1.0]{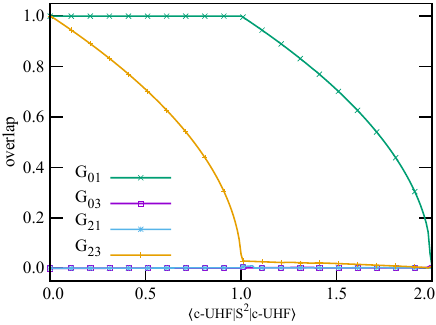}
  \caption{} 
  \label{fig:LiH_500_MOoverlap}
\end{subfigure}
\caption{For LiH/6-31G: Comparison of c-UHF, NOCI(2), NOCI(6), and FCI energies across the full range of imposed constraints $\langle\textrm{c-UHF}|S^2|\textrm{c-UHF}\rangle$ at (a) $R=2.75\,a_{0}$ and (c) $R=5.0\,a_{0}$. The overlap between the $\alpha$ (0, 2) and $\beta$ (1, 3) spin-orbitals at (b) $R=2.75\,a_{0}$ and (d) $R=5.00\,a_{0}$.}
\label{fig:LiH_631g}
\end{figure}

Figures~\ref{fig:LiH_275_NOCI} and \ref{fig:LiH_500_NOCI} illustrate the c-UHF and NOCI($k$) energies as functions of the imposed $\langle \textrm{c-UHF}|S^2|\textrm{c-UHF} \rangle$ value at $R=2.75\,a_{0}$, and $R=5.00\,a_{0}$. 
Additionally, Figures~\ref{fig:LiH_275_MOoverlap} and \ref{fig:LiH_500_MOoverlap} depict the overlap between high-spin and low-spin orbitals during the symmetry-breaking process for both bond lengths. Initially, spin-breaking occurs only in the valence space, with the core electrons of $\ce{Li}$ participating only for $\langle\textrm{c-UHF}|S^2|\textrm{c-UHF}\rangle \geq 1$. Consequently, the NOCI($k$) calculation that fully restores spin symmetry is equivalent to NOCI(2) in the interval $\langle \textrm{c-UHF} | S^2 | \textrm{c-UHF} \rangle \in [0,1]$, featuring a variational minimum at a non-zero value of $\langle \textrm{c-UHF} | S^2 | \textrm{c-UHF} \rangle$. However, NOCI(6) is required in the interval $[1,2]$ to achieve full symmetry restoration. 

Although $\ce{LiH}$ is a $4$-electron system, the variational minimum of SPcUHF is obtained at the NOCI(2) level, where only the valance electrons are involved in the symmetry breaking.
Overall, SPcUHF can predict 33.68\,\% and 81.25\,\% of the correlation energy at bond lengths  $R=2.75\,a_{0}$ and $R=5.00\,a_{0}$, respectively, with the same computational cost as NOCI(2), demonstrating its effectiveness in capturing static correlation with a reasonable computational efficiency. Additional dynamical correlation can be incorporated using the second-order perturbative correction NOCI(2)+PT2, predicting 81.69\,\% and 94.78\,\% of the correlation, respectively. This inclusion of dynamic correlation  brings the SPcUHF energies much closer to the threshold of chemical accuracy.



\begin{figure}[htb]
\centering
\begin{subfigure}{.5\textwidth}
  \centering
  \includegraphics[scale=1.0]{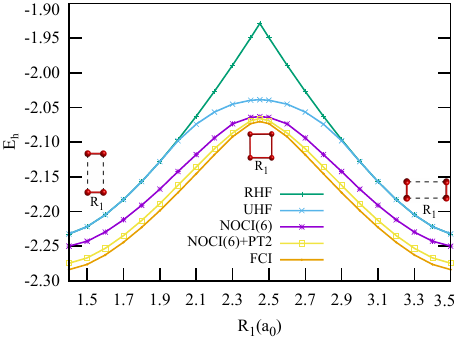}
  \caption{}
  \label{fig:H4_noci}
\end{subfigure}%
\begin{subfigure}{.5\textwidth}
  \centering
  \includegraphics[scale=1.0]{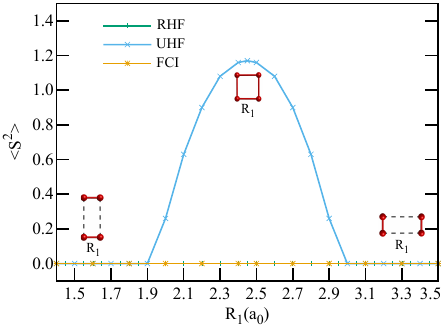}
  \caption{}
  \label{fig:H4_expSS}
\end{subfigure}
\caption{(a) Total energy and (b) $\expval*{\hat{S}^{2}}$ value for the RHF and UHF, NOCI(6), NOCI(6)+PT2, and FCI ground states along the \ce{H4}/6-31G binding curve.}
\label{fig:H4_631g}
\end{figure}

The Paldus \ce{H_4} transition from the rectangular $\mathrm{D_{2h}}$ to the square $\mathrm{D_{4h}}$ symmetry is a typical example of a multi-reference system in which a transition is observed from a double singlet bond to an anti-aromatic compound. \cite{paldus1970stability} Initially, two \(\ce{H2}\) molecules with a internal bond length of \(R_{1}=1.4\,a_{0}\) are separated by a distance of \(R_{2}=3.5\,a_{0}\), forming a rectangular shape. By increasing \(R_{1}\) and decreasing \(R_{2}\), the \(\ce{H4}\) molecule is transformed into a square configuration where \(R_{1}=R_{2}=2.45\,a_{0}\). Further stretching \(R_{1}\) and shortening \(R_{2}\) results in a rectangular shape, equivalent to the original conformation rotated by $90^{\circ}$.
Figures~\ref{fig:H4_noci} and \ref{fig:H4_expSS} show the RHF and UHF energies and $\langle S^2\rangle$ values as functions of $R_{1}$, illustrating the spontaneous spin symmetry-breaking near the square $\mathrm{D_{4h}}$ configuration. UHF lowers the energy compared to RHF by breaking spin symmetry in the range $R_{1} \in 2.0$ to $2.9\,a_{0}$. Notably, unlike in \ce{LiH} where $\langle\textrm{c-UHF}| S^2||\textrm{c-UHF}\rangle \leq 1$ at the variational minimum, the $\langle\textrm{c-UHF}|S^2|\textrm{c-UHF}\rangle$ value reaches 1.17 at the $\mathrm{D_{4h}}$ configuration ($R_{1}=2.45\,a_{0}$) as all four electrons become unpaired and contribute to the anti-aromatic character of the system.

\begin{figure}[htb]
\centering
\begin{subfigure}{.5\textwidth}
  \centering
  \includegraphics[scale=1.0]{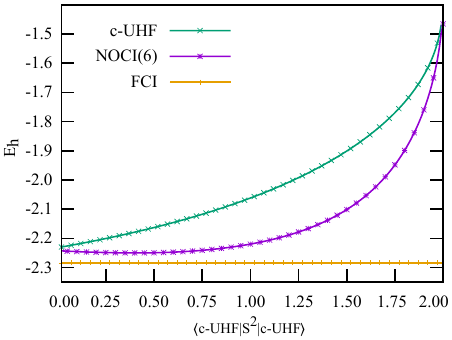}
  \caption{}
  \label{fig:H4_14_NOCI}
\end{subfigure}%
\begin{subfigure}{.5\textwidth}
  \centering
  \includegraphics[scale=1.0]{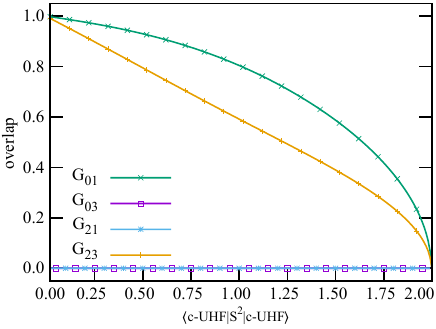}
  \caption{}
  \label{fig:H4_14_overlap}
\end{subfigure}
  \begin{subfigure}{.5\textwidth}
  \centering
  \includegraphics[scale=1.0]{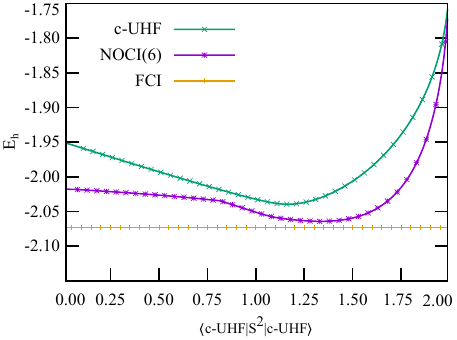}
  \caption{}
  \label{fig:H4_25_NOCI}
\end{subfigure}%
\begin{subfigure}{.5\textwidth}
  \centering
  \includegraphics[scale=1.0]{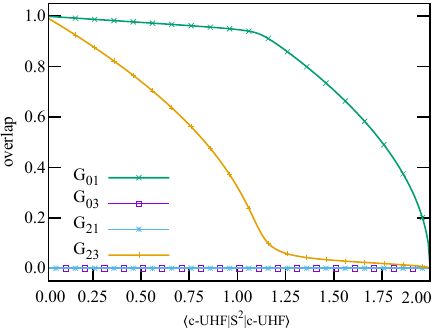}
  \caption{}
  \label{fig:H4_25_overlap}
\end{subfigure}
\caption{Comparison of c-UHF, NOCI(2), NOCI(6), and FCI energies across the full range of imposed constraints $\langle S^2\rangle$ for For \ce{H_4}/6-31G at (a) $R_{1}=1.4\,a_{0}$ and (c) $R_{1}=2.5\,a_{0}$. The overlap between the $\alpha$ (0, 2) and $\beta$ (1, 3) molecular spin-orbitals at (b) $R_{1}=1.4\,a_{0}$ and (d) $R_{1}=2.5\,a_{0}$.}
\label{fig:H4_14&25}
\end{figure}

The SPcUHF has been applied to \ce{H4} throughout this transition. In Figure~\ref{fig:H4_14&25}, results are shown for the full range of $\langle\textrm{c-UHF}|S^2|\textrm{c-UHF}\rangle$ at two bond lengths: (1) $R_{1}=1.4\,a_{0}$, where spin symmetry is largely preserved, and (2) $R_{1}=2.50\,a_{0}$, within the spontaneous symmetry-breaking regime.
The chemical picture of the $\mathrm{D_{2h}}$ configuration comprises two individual and largely independent \ce{H_2} molecules.  As both \ce{H_2} are approximately closed shell at $R_1 = 1.4\,a_0$, the amount of static correlation is expected to be limited.  The absence of a derivative discontinuity at $\langle\textrm{c-UHF}|S^2|\textrm{c-UHF}\rangle = 1$ hints that all four electrons are involved in the symmetry breaking from the start, which is confirmed by the overlap matrix between the spin-up and spin-down orbitals in Figures~\ref{fig:H4_14_overlap} and \ref{fig:H4_25_overlap}. Hence, NOCI(6) fully restores the spin symmetry, with a energy minimum in the range $\langle\textrm{c-UHF}|S^2|\textrm{c-UHF}\rangle \in [0,1]$.

 A qualitatively different behavior is observed closer to the square \ce{H4} configuration. The non-interacting molecular orbital picture of \ce{H_4} in the square $\mathrm{D_{4h}}$ configuration involves two electrons in an orbital that transforms as the low-lying $\mathrm{A_{1g}}$ trivial representation and the other two electrons in energetically higher orbitals that transform as the two-fold degenerate $\mathrm{E_u}$ irreducible representation.  For this reason, we observe that the onset of symmetry breaking only happens within the $\mathrm{E_u}$ (valence) electrons.  To arrive at the multi-reference valence-bond resonance structure of the square \ce{H_4}, the symmetry needs to be broken at the level of all four electrons, with a variational minimum for NOCI(6) in the $\langle\textrm{c-UHF}|S^2|\textrm{c-UHF}\rangle \in [1,2]$ range.

Figure~\ref{fig:H4_noci} illustrates the NOCI(6) energy surface across the full rectangle-square-rectangle rearrangement.  Although NOCI(6) only predicts a modest 34.15\,\% of the total correlation energy at $R_1=1.40\,a_0$, it is remarkably effective at capturing static correlation around the resonance regime, capturing 94.98\,\% of the correlation energy at $R_{2}=2.45\,a_{0}$. Including dynamic correlation using the NOCI+PT2 correction increases the accuracy to 81.99\,\% and 96.98\,\% correlation energy at the two geometries, respectively.

\subsection{6 and 8-electron systems: \ce{C}/6-31G and \ce{Be2}/6-31G}\label{subsection:6e}

\begin{figure}[htb]
\centering
\begin{subfigure}{.5\textwidth}
  \centering
  \includegraphics[scale=1.0]{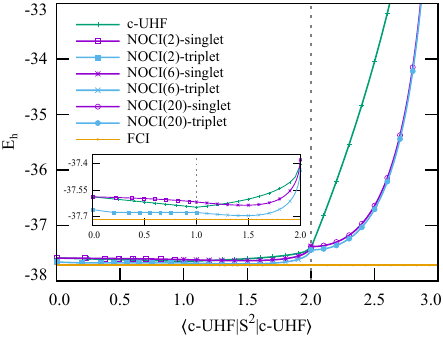}
  \caption{}
  \label{fig:C_631g_NOCI}
\end{subfigure}%
\begin{subfigure}{.5\textwidth}
  \centering
  \includegraphics[scale=1.0]{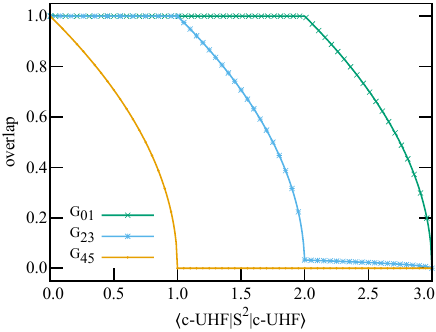}
  \caption{}
  \label{fig:C_631g_overlap}
\end{subfigure}
\caption{For C/6-31G over the full range of imposed constraint $\langle\textrm{c-UHF}|S^2|\textrm{c-UHF}\rangle$: (a) The c-UHF, NOCI(2), NOCI(6), and NOCI(20) energies of the singlet and triplet states and (b) The overlap between $\alpha$ (0, 2, 4) and $\beta$ (1, 3, 5) molecular spin-orbitals.}
\label{fig:C_631g}
\end{figure}

For the six-electron \ce{C} atom, unpairing $M=2,4$ or $6$ electrons will lead to NOCI(2), NOCI(6) and NOCI(20) calculations respectively, based on the $\binom{M}{M/2}$ binomial formula. 
Figure~\ref{fig:C_631g_overlap} presents the overlap matrix elements $G_{2i, 2j+1}$ of the \ce{C} atom in c-UHF as a function of the imposed $\langle\textrm{c-UHF}|S^2|\textrm{c-UHF}\rangle$ constraint.
For the C atom, the electrons in the 2p valence orbitals are the first to break spin symmetry in the $\langle\textrm{c-UHF}|S^2|\textrm{c-UHF}\rangle \in [0,1]$ regime, followed by the 2s and 1s core electrons in the $[1,2]$ and $[2,3]$ ranges, respectively. An interesting aspect of NOCI($k$) computations is that they provide projections into all possible spin spaces, including singlet, triplet, and quintuplet states for C.
The SPcUHF results in Figure~\ref{fig:C_631g_NOCI} show that the triplet state is lower in energy compared to the singlet state, which is expected based on Hund's rules for the electronic structure of atoms. NOCI(20) provides no further improvement over NOCI(6), showing that most of the static correlation in \ce{C} can be captured by breaking spin symmetry of the $2$s and $2$p valence electrons. SPcUHF captures 61.92\% of the total correlation energy in the \ce{C} atom, primarily attributable to static correlation. This value increases to 97.27\% when using NOCI(6)+PT2, indicating that most of the electron correlation is accounted for. For comparison, the CASSCF(6,(2,2)) method captures 91.91\% of the total correlation energy.
\\

\begin{figure}[htb]
\centering
\begin{subfigure}{.5\textwidth}
  \centering
  \includegraphics[scale=1.0]{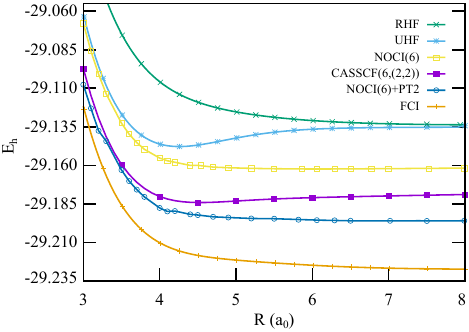}
  \caption{}
  \label{fig:Be2_RUHFpes}
\end{subfigure}%
\begin{subfigure}{.5\textwidth}
  \centering
  \includegraphics[scale=1.0]{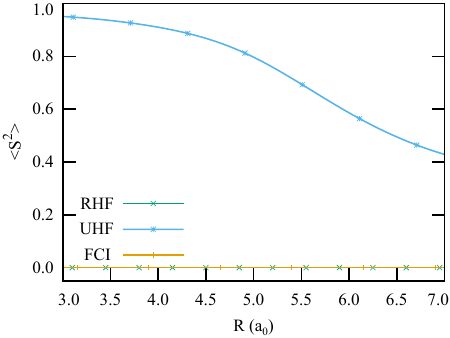}
  \caption{}
  \label{fig:Be2_RUHFspin}
\end{subfigure}
\caption{(a) Total energy and (b) $\expval*{\hat{S}^{2}}$ value of the RHF, UHF, and FCI ground states for the \ce{Be_2}/6-31G binding curve.}
\label{fig:Be2_631g_ruhf}
\end{figure}

Lastly, the \ce{Be2} diatomic molecule is selected to further evaluate the SPcUHF method in larger systems. For an eight-electron system, if all spin-orbitals are unpaired \(M=8\), according to the \(\binom{M}{M/2}\) binomial formula, then a NOCI(70) calculation can fully restore the spin symmetry. However, smaller NOCI($k$) calculations may provide sufficient accuracy at lower computational cost.  Figure~\ref{fig:Be2_631g_ruhf} compares the ground-state potential energy surfaces for RHF, UHF, SPcUHF, SPcUHF+PT2, with CASSCF(6,(2,2)) and FCI.  Although UHF systematically improves the energy compared to RHF over all bond distances, it enormously underestimates the equilibrium bond distance with $R_{\textrm{UHF}} = 4.0\,a_0$.  For the SPcUHF analysis, we consider the two bond lengths \(R = 4.00\,a_0\), and \(R = 5.75\,a_0\).

\begin{figure}[htb]
\centering
\begin{subfigure}{.5\textwidth}
  \centering
  \includegraphics[scale=1.0]{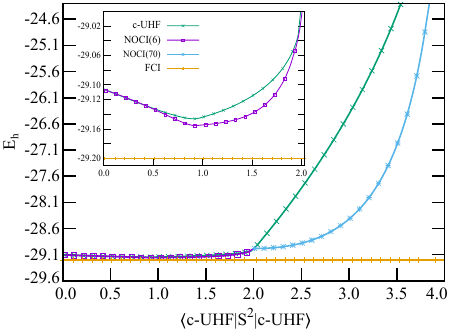}
  \caption{}
  \label{fig:Be2_400_NOCI}
\end{subfigure}%
\begin{subfigure}{.5\textwidth}
  \centering
  \includegraphics[scale=1.0]{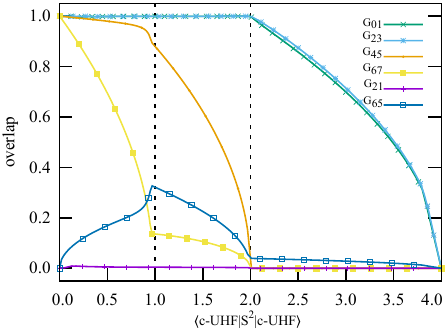}
  \caption{}
  \label{fig:Be2_400_overlap}
\end{subfigure}
  \begin{subfigure}{.5\textwidth}
  \centering
  \includegraphics[scale=1.0]{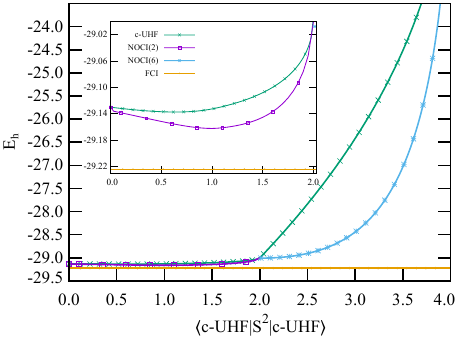}
  \caption{}
  \label{fig:Be2_575_NOCI}
\end{subfigure}%
\begin{subfigure}{.5\textwidth}
  \centering
  \includegraphics[scale=1.0]{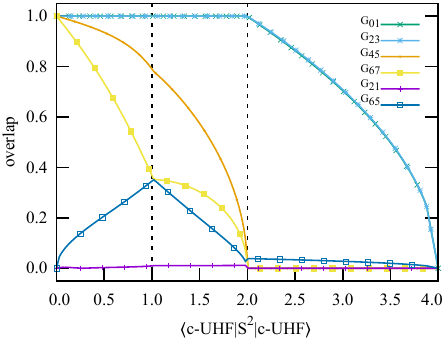}
  \caption{}
  \label{fig:Be2_575_overlap}
\end{subfigure}
\caption{Comparison of the c-UHF, NOCI(6), NOCI(70) and FCI ground-state energies for \ce{Be_2}/6-31G across the full range of imposed constraints $\langle S^2\rangle$ at (a) $R=4.00\,a_{0}$ and (c)$R=5.75\,a_{0}$. The overlap between the $\alpha$ (0, 2, 4, 6) and $\beta$ (1, 3, 5, 7) molecular spin-orbitals at (b) $R=4.00\,a_{0}$ and (d) $R=5.75\,a_{0}$.}
\label{fig:Be2_400&575}
\end{figure}

The overlap matrix elements, $G_{2i , 2j+1}$, for the $\alpha_i$ (0, 2, 4, 6) and $\beta_j$ (1, 3, 5, 7) spin-orbitals as a function of the imposed $\langle\textrm{c-UHF}|S^2|\textrm{c-UHF}\rangle$ constraint at both bond lengths are depicted in Figures~\ref{fig:Be2_400_overlap} and \ref{fig:Be2_575_overlap}. The symmetry-breaking process begins with the unpairing of the two highest-energy valence spin-orbital pairs (6 \& 7 and 4 \& 5) within the $[0,2]$ region. Therefore, we see that NOCI(6) is sufficient in this interval and provides comparable accuracy to NOCI(70).  After $\langle\textrm{c-UHF}|S^2|\textrm{c-UHF}\rangle = 2$, when the two highest-energy  valence pairs are completely symmetry broken and separated, the first two core electron pairs (0 \& 1 and 2 \& 3) begin to unpair simultaneously.  NOCI(70) is therefore required to fully restore the symmetry in this range.  NOCI(20) results in the $[0,2]$ interval reduce to NOCI(6), whereas in the $[2,4]$ interval, they represent an incomplete spin-projected wave function, as they cannot fully restore spin symmetry due to some unpaired (core) spin orbitals missing in the projection.

The SPcUHF analysis for both bond lengths are plotted in Figures~\ref{fig:Be2_400_NOCI} and \ref{fig:Be2_575_NOCI}. NOCI(6) captures 47.26\% of the correlation energy at $4.00,a_0$ and 34.34\% at $5.75,a_0$. NOCI(70) does not provide additional improvement, as breaking the core pair of electrons is not energetically favorable. The SPcUHF algorithm offers the advantage of capturing a significant portion of the static correlation while reducing the total computational cost from NOCI(70) to NOCI(6). These results can be further improved by including additional dynamic correlation through NOCI(6)+PT2, which captures 78.76\% and 69.32\% of the correlation energy at the two bond lengths, respectively.
For these bond lengths, the SPcUHF and SPcUHF+PT2 results are listed in Table~\ref{table:Be2comparison}, alongside the CASSCF(6,(2,2)) and FCI results, providing a comprehensive assessment of our method's ability to capture the missing correlation.

\begin{table}[H]
\caption{Total energy and the percentage of the correlation energy captured in RHF, SPcUHF, SPcUHF+PT2, CASSCF, and Full-CI calculations on \ce{Be2}/6-31G.}
\label{table:Be2comparison}
\begin{ruledtabular}
 \begin{tabular}{lcc} 
 & \multicolumn{1}{c}{$R = \SI{4.00}{a_{0}}$}
\\
 \cline{2-2} 
 & $E_h$  & $\,\%\ \text{Corr.}$
\\ \hline
SPcUHF & -29.15513 & 47.26
\\
CASSCF(6,(2,2)) & -29.18025 & 71.29 
\\
SPcUHF+PT2 & -29.18805 & 78.76
\\
FCI & -29.21025 & 100.00
\\ \hline \hline
 & \multicolumn{1}{c}{$R = \SI{5.75}{a_{0}}$}
\\
 \cline{2-2} 
& $E_h$  & $\,\%\ \text{Corr.}$
\\ \hline
SPcUHF & -29.16235 & 34.34 
\\
CASSCF(6,(2,2)) & -29.18139 & 54.57
\\ 
SPcUHF+PT2 & -29.19527 & 69.32 
\\

FCI & -29.22415 & 100.00
 \end{tabular}
\end{ruledtabular}
\end{table}

\subsection{Excited States}

\begin{figure}[H]
\centering
\begin{subfigure}{.5\textwidth}
  \centering
  \includegraphics[scale=1.0]{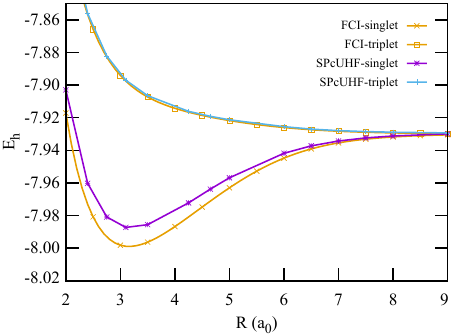}
  \caption{}
  \label{fig:LiH_es}
\end{subfigure}%
\begin{subfigure}{.5\textwidth}
  \centering
  \includegraphics[scale=1.0]{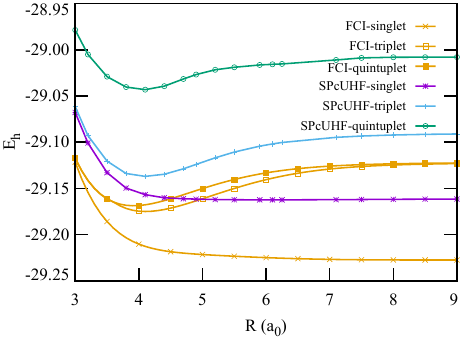}
  \caption{}
  \label{fig:Be2_es}
\end{subfigure}
\caption{Ground and excited states obtained by SPcUHF method compared to FCI for (a) \ce{LiH}/6-31G and (b) \ce{Be2}/6-31G systems.}
\label{fig:LiH_Be2_ES}
\end{figure}

An added advantage of the NOCI approach is that, as a CI approach, excited states come at the same computational cost as ground states, depending on the diagonalization procedure.  This is a relevant observation within the context of, for instance, photochemistry.  To assess the capacity of the SPcUHF method to produce excited states, we present the lowest energy states within each spin sector for the \ce{LiH} and \ce{Be2} dimers in Figures \ref{fig:LiH_es} and \ref{fig:Be2_es} respectively.  For \ce{LiH}, SPcUHF delivers a triplet state energy that is very close to exact FCI results, demonstrating its effectiveness in capturing excited states dominated by static correlation.  We also observe that the triplet and singlet state converge to the same energy at infinite bond distance.  In the case of \ce{Be2}, SPcUHF energies for the triplet state deviate by approximately $0.040\,\unit{E_h}$ from FCI, with a parallellity error of $0.024\,\unit{E_h}$.  For the quintuplet state, the energies deviate by approximately $0.120\,\unit{E_h}$, with a non-parallellity error of $0.024\,\unit{E_h}$.

\section{Conclusion}\label{sec:conclusions}

We present the spin-projected constrained-unrestricted Hartree--Fock (SPcUHF) method, an approach based on symmetry breaking and restoration at the mean-field HF level. This method breaks the spin symmetry with constrained-UHF and then restores the broken spin symmetry by forming a NOCI basis using all possible configurations generated from a Clebsch--Gordon recoupling scheme in spin space. Our analysis reveals that the spin symmetry-breaking process in c-UHF occurs sequentially, with electron pairs unpairing one at a time, beginning with the highest-energy valence electron pair. At the RHF level, all spin-up and spin-down electrons are paired in identical spatial orbitals. As symmetry breaking initiates, the highest-energy occupied pair of electrons unpair first, meaning the spatial orbitals for spin-up and spin-down electrons in this pair become distinct and nonorthogonal. This unpairing process continues sequentially through the remaining pairs, reaching down to the core pairs. Given that static correlation, which is closely associated with bond breaking or competing resonance structures, predominantly arises in the valence electrons, the global energy minimum is typically found before all pairs are unpaired. Consequently, the number of projected states required in the NOCI basis for full symmetry restoration can be reduced without sacrificing accuracy. We demonstrate the efficacy of SPcUHF on systems with 4, 6, and 8 electrons, where both static and dynamic correlations are significant, showing that our method captures these effects efficiently and accurately.

\section*{Acknowledgements}

HGAB gratefully thanks Downing College, Cambridge for financial support through the Kim and Julianna Silverman Research Fellowship. SDB acknowledges the NSERC Canada Research Chair and Discovery Grant program for financial support.

\bibliography{spinProjection}

\clearpage
\appendix

\section{spin-properties of four-electron systems}\label{section:appendix}

A constructive proof that the NOCI($k$) eigenstates are fully spin-projected would involve the construction of general Clebsch-Gordon (CG) recoupled spin configurations, which is beyond the scope of this work.  Nevertheless, the explicit construction for the NOCI($6$) four-electron case is insightful for our purpose.  We will show here that the CG recoupling coeficients for four electrons bring the NOCI overlap matrix between the six configurations (\ref{spin:noci:6states}) into block-diagonal shape with respect to their total spin sectors $S=(0,1,2)$, with degeneracies (2,3,1) respectively.  We will use the bar-notation $\{|i\rangle=a_{i\su}^\dag|\theta\rangle,|\bar{i}\rangle=a_{i\sd}^\dag|\theta\rangle\}$ and notation introduced in equation (\ref{spin:proj:4electrons})
\begin{equation}
|(S_{01}S_{23})S\rangle = [[a_0^\dag a_1^\dag]^{S_{01}}[a_2^\dag a_3^\dag]^{S_{23}}]^{S}_0|\theta\rangle
\end{equation}
for the uncoupled and recoupled states respectively to simplify the notation.  The unitary transformation $U_{R}$ that brings the uncoupled (\ref{spin:noci:6states}) to the recoupled set of states is given by
\begin{equation}
    \left(\begin{array}{c} 
    {}|(00)0\rangle \\ 
    {}|(11)0\rangle \\ 
    {}|(01)1\rangle \\ 
    {}|(10)1\rangle \\ 
    {}|(11)1\rangle \\ 
    {}|(11)2\rangle \\
    \end{array}\right)
    =\left(\begin{array}{cccccc}
     0 & \frac{1}{2} &  -\frac{1}{2} & -\frac{1}{2} & \frac{1}{2} & 0 \\
     \frac{\sqrt{3}}{3} & -\frac{\sqrt{3}}{6} & -\frac{\sqrt{3}}{6} & -\frac{\sqrt{3}}{6} & -\frac{\sqrt{3}}{6} &  \frac{\sqrt{3}}{3} \\
     0 & \frac{1}{2} &  -\frac{1}{2} & \frac{1}{2} & -\frac{1}{2} & 0 \\
     0 & \frac{1}{2} & \frac{1}{2} &  -\frac{1}{2} & -\frac{1}{2} & 0 \\
     \frac{\sqrt{2}}{2} & 0 & 0 & 0 & 0 & -\frac{\sqrt{2}}{2} \\
     \frac{\sqrt{6}}{6} & \frac{\sqrt{6}}{6} & \frac{\sqrt{6}}{6} & \frac{\sqrt{6}}{6} & \frac{\sqrt{6}}{6} &  \frac{\sqrt{6}}{6} \\
     \end{array}\right)
    \left(\begin{array}{c} 
    |01\bar{2}\bar{3}\rangle \\
    |0\bar{1}2\bar{3}\rangle \\
    |\bar{0}12\bar{3}\rangle \\
    |0\bar{1}\bar{2}3\rangle \\
    |\bar{0}1\bar{2}3\rangle \\
    |\bar{0}\bar{1}23\rangle 
    \end{array}\right)
\end{equation}\label{appendix:cgtransfo}
To fix ideas, the $|0\bar{1}2\bar{3}\rangle$ uncoupled state would be the one arising from a four-electron c-UHF computation, with the even $(0,2)$ and odd $(1,3)$ orbital indices referring to the spin-up and -down electrons respectively.  The overlap matrix between the four spatial orbitals involved is therefore given by
\begin{equation}
\mathbf{G} = \left(\begin{array}{cccc}
     1 & G_{01} &      0 & G_{03} \\
G_{10} &      1 & G_{12} &      0 \\
     0 & G_{21} &      1 & G_{23} \\
G_{30} &      0 & G_{32} &      1\end{array}\right)
\end{equation}
in which the notation introduced in the anticommutation relations (\ref{spin:tensor:updown}) has been used.  From this, the NOCI(6) overlap matrix, called $M$ in the present section, constructed from the uncoupled states is given by
\begin{equation}
\mathbf{M}=\left(\begin{array}{cccccc}
 f_1 & -G_{21}^{2} & -f_3f_4 & -f_3f_4 &  -G_{03}^{2} & f_4^2 \\
 -G_{21}^{2}& 1 & -G_{01}^{2} & -G_{23}^{2}  & (f_3-f_4)^2 & -G_{03}^{2}\\
 -f_3f_4 & -G_{01}^{2} & f_2 & f_3^2 & -G_{23}^{2}  & -f_3f_4\\
 -f_3f_4 & -G_{23}^{2} &f_3^2 & f_2  & -G_{01}^{2} & -f_3f_4\\
 -G_{03}^{2} & (f_3-f_4)^2 & -G_{23}^{2} & -G_{01}^{2}  & 1 & -G_{21}^{2}\\
 f_4^2 & -G_{03}^{2} & -f_3f_4 & -f_3f_4 & -G_{21}^{2} & f_1 \\
\end{array}\right)\label{eq:appendix:m}
\end{equation}
with the $f_i$ functions defined as 
\begin{align}
    f_1 & = (1-G_{01}^2)(1-G_{23}^2)\\
    f_2 & = (1-G_{03}^2)(1-G_{21}^2)\\
    f_3 & = G_{01}G_{23}\\
    f_4 & = G_{03}G_{21}
\end{align}
Obviously, the NOCI(6) overlap matrix reduces to the unit matrix whenever all non-diagonal overlaps $G_{2i,2j+1}$ vanish between the spatial orbitals in the spin-up and -down sector.  In the general case, the matrix $M$ reduces to a block diagonal shape when employing the unitary recoupling self-similarity transformation
\begin{equation}
U_RMU_{R}^T=   \left(\begin{array}{cc|ccc|c} 
    {}M^{(0)}_{11} & {}M^{(0)}_{12} & \cdot & \cdot & \cdot & \cdot  \\
    {}M^{(0)}_{21} & {}M^{(0)}_{22} & \cdot & \cdot & \cdot & \cdot  \\
    \hline
    \cdot & \cdot & M^{(1)}_{11} & M^{(1)}_{12} & M^{(1)}_{13} & \cdot  \\
    \cdot & \cdot & M^{(1)}_{21} & M^{(1)}_{22} & M^{(1)}_{23} & \cdot  \\
    \cdot & \cdot & M^{(1)}_{31} & M^{(1)}_{32} & M^{(1)}_{33} & \cdot  \\
    \hline
    \cdot & \cdot & \cdot & \cdot & \cdot & M^{(2)}_{11}\end{array}\right)
\end{equation}
with the $M^{(S)}$ block matrices referring to their respective recoupled spin $S$ spaces in (\ref{spin:proj:4electrons}).  For completeness, the matrix elements of the block matrices are given by 
\begin{align}
M^{(0)}_{11}&=1+G_{01}^2+G_{23}^2-\tfrac{1}{2}G_{03}^2-\tfrac{1}{2}G_{21}^2+f_3^2-f_3f_4+f_4^2\\
M^{(0)}_{12}&=M^{(0)}_{21}=-\tfrac{\sqrt{3}}{2}[G_{03}^2+G_{21}^2-2f_3f_4]\\
M^{(0)}_{22}&=1-G_{01}^2-G_{23}^2+\tfrac{1}{2}G_{03}^2+\tfrac{1}{2}G_{21}^2+f_3^2+f_3f_4+f_4^2
\end{align}
\begin{align}
M^{(1)}_{11}&=1+G_{01}^2-G_{23}^2-\tfrac{1}{2}G_{03}^2-\tfrac{1}{2}G_{21}^2-f_3^2+f_3f_4\\
M^{(1)}_{12}&=M^{(1)}_{21}=\tfrac{1}{2}{G}_{03}^2+\tfrac{1}{2}G_{21}^2+f_3f_4-f_4^2\\
M^{(1)}_{22}&=1-G_{01}^2+G_{23}^2-\tfrac{1}{2}G_{03}^2-\tfrac{1}{2}G_{21}-f_3^2+f_3f_4\\
M^{(1)}_{13}&=M^{(1)}_{23}=M^{(1)}_{31}=M^{(1)}_{32}=\tfrac{\sqrt{2}}{2}[G_{03}^2-G_{21}^2]\\
M^{(1)}_{33}&=1-G_{01}^2-G_{23}^2+f_3^2-f_4^2
\end{align}
\begin{equation}
M^{(2)}_{11}=1-G_{01}^2-G_{23}^2-G_{03}^2-G_{12}^2+(f_3-f_4)^2
\end{equation}
This result confirms that the NOCI(6) basis constructed from all spin configurations with four electrons in spatial orbitals generated from a c-UHF computation is indeed compatible with the standard Clebsch-Gordon spin recouplings into $S=0,1,2$ sectors.  

The situation becomes more interesting when considering the situation in which the spatial symmetry is only broken in one spatial orbital pair.   To fix ideas, let's assume that the spatial orbitals 0 and 1 are equivalent, and spatial orbitals 2 and 3 are different.  This is the picture present in the middle panel of Figure \ref{fig:bepairbreaking}.  In this situation, we have
\begin{equation}
    G_{01}=1,\quad\textrm{\&}\quad G_{03}=G_{21}=0,
\end{equation}
with cross-overlaps $G_{03}=G_{21}=0$ occurring because spatial orbitals 0 and 1 are now equivalent.  As a consequence, all $M^{(S)}$ matrix elements vanish for all $S$, except
\begin{align}
M^{(0)}_{11} &=2+2G_{23}^2 \\
M^{(1)}_{11} &=2-2G_{23}^2, 
\end{align}
meaning that the $S=0$ and $S=1$ spaces reduce to only one non-vanishing state, and the $S=2$ state disappears completely.  This is to be expected as the problem basically reduced to a closed singlet in the (01) orbitals in the presence of two symmetry-broken electrons that can couple to an open singlet or triplet, but not a quintuplet. 

Finally, in the limit where also the spatial orbitals 2 and 3 are the same, we additionally have 
\begin{equation}
    G_{23}=1,
\end{equation}
and we retain the RHF picture at the left of Figure \ref{fig:bepairbreaking}.  In this case, the only surviving matrix element in the NOCI(6) overlap matrix is 
\begin{equation}
    M^{(0)}_{11} =4
\end{equation}
and also the triplet state has disappeared entirely.

\end{document}